\documentclass[10pt,letterpaper]{article}
\usepackage[top=0.85in,left=2.75in,footskip=0.75in]{geometry}

\usepackage{changepage}

\usepackage[utf8]{inputenc}

\usepackage{color}

\usepackage{textcomp,marvosym}

\usepackage{fixltx2e}

\usepackage{amsmath,amssymb}

\usepackage{cite}

\usepackage{nameref,hyperref}

\usepackage{microtype}
\DisableLigatures[f]{encoding = *, family = * }

\usepackage{rotating}

\raggedright
\usepackage[aboveskip=1pt,labelfont=bf,labelsep=period,justification=raggedright,singlelinecheck=off]{caption}

\makeatletter
\renewcommand{\@biblabel}[1]{\quad#1.}
\makeatother

\date{}

\usepackage{lastpage,fancyhdr,graphicx}
\fancyheadoffset[L]{2.25in}
\fancyfootoffset[L]{2.25in}
\newcommand{\la}{\langle}
\newcommand{\ra}{\rangle}

\begin{document}
\vspace*{0.35in}

% Title must be 150 characters or less
\begin{flushleft}
{\Large
\textbf\newline{An Analysis of the Matching Hypothesis in Networks}
}
\newline
% Insert Author names, affiliations and corresponding author email.
\\
Tao Jia\textsuperscript{1,2,3*},
Robert F. Spivey\textsuperscript{3,4},
Boleslaw Szymanski\textsuperscript{1,2,5},
Gyorgy Korniss\textsuperscript{1,3}
\\
\bf{1} Social Cognitive Networks Academic Research Center, Rensselaer Polytechnic Institute, Troy, NY, 12180 USA
\\
\bf{2} Department of Computer Science, Rensselaer Polytechnic Institute, Troy, NY, 12180 USA
\\
\bf{3} Department of Physics, Applied Physics and Astronomy, Rensselaer Polytechnic Institute, Troy, NY, 12180 USA
\\
\bf{4} Department of Electrical and Computer Engineering, Duke University, Durham, NC, 27708 USA
\\
\bf{5} Spo\l{}eczna Akademia Nauk, \L{}\'{o}d\'{z}, Poland
\\
% Insert additional author notes using the symbols described below. Insert symbol callouts after author names as necessary.
% 
% Remove or comment out the author notes below if they aren't used.
%
% Primary Equal Contribution Note
%\Yinyang These authors contributed equally to this work.

% Additional Equal Contribution Note
%\ddag These authors also contributed equally to this work.

% Current address notes
%\textcurrency a Insert current address of first author with an address update
% \textcurrency b Insert current address of second author with an address update
% \textcurrency c Insert current address of third author with an address update

% Deceased author note
%\dag Deceased

% Group/Consortium Author Note
%\textpilcrow Insert Collaborative Author line here

* E-mail: taojia82@gmail.com
\end{flushleft}
% Please keep the abstract below 300 words
\section*{Abstract}
The matching hypothesis in social psychology claims that people are
more likely to form a committed relationship with someone equally
attractive. Previous works on stochastic models of human mate choice
process indicate that patterns supporting the matching hypothesis
could occur even when similarity is not the primary consideration in
seeking partners. Yet, most if not all of these works concentrate on
fully-connected systems. Here we extend the analysis to networks.
Our results indicate that the correlation of the couple's
attractiveness grows monotonically with the increased average
degree and decreased degree diversity of the network. This
correlation is lower in sparse networks than in
fully-connected systems, because in the former less attractive
individuals who find partners are likely to be coupled with ones who
are more attractive than them. The chance of failing to be matched
decreases exponentially with both the attractiveness and the degree.
The matching hypothesis may not hold when the degree-attractiveness
correlation is present, which can give rise to negative attractiveness correlation. Finally, we find that the ratio
between the number of matched couples and the size of the maximum
matching varies non-monotonically with the average degree of the
network. Our results reveal the role of network topology in the process of human mate choice and bring insights into future investigations of different matching processes in networks.

% Please keep the Author Summary between 150 and 200 words
% Use first person. PLOS ONE authors please skip this step. 
% Author Summary not valid for PLOS ONE submissions.   
%\section*{Author Summary}

%\linenumbers

\section*{Introduction}
The process of pairing and matching between members of two disjoint
groups is ubiquitous in our society. The underlying mechanism can be
purely random, but in general decisions on selections are guided by
rational choices, such as the relationship between advisor and
advisee, the employment between employer and employee and the
marriage between heterosexual male and female individuals. In many of these
cases, similarities between the two paired parties are widely
observed, such as similar research interests between the advisor and
advisee and matched market competitiveness between the executives
and the company. The principle of homophily, the tendency of
individuals to associate and bond with others who are similar to
them, can be applied to explain such similarities
\cite{mcpherson2001birds}. Yet, in some cases different mechanisms
may be at work in addition to simply seeking similarities. For
example, it has been discovered that people end up in committed
relationship in which partners are likely to be of similar
attractiveness, as predicted by the matching hypothesis in the field
of social psychology \cite{walster1966importance,
berscheid1971physical}. However, if the closeness in attractiveness
is the goal when searching for partners, one needs an objective
self-estimation of it, which is rarely the case
\cite{taylor2011out}. Furthermore, it is found in social experiments
that people tend to pursue or accept highly desirable individuals
regardless of their own attractiveness \cite{berscheid1971physical,
taylor2011out}. These findings suggest that the observed
similarities may not be solely caused by explicitly seeking
similarities. In some previous works, stochastic models are applied
to simulate the process of human mate choice
\cite{kalick1986matching, alpern1999strategic, alpern2005strategic,
simao2003emergent, ramsey2011mutual, smaldino2012human}. By simply
assuming that highly attractive individuals are more likely to be
accepted, the system generates patterns supporting the matching
hypothesis even when similarity is not directly considered in the
partner selection process \cite{kalick1986matching}. Nevertheless,
most if not all of these works (with a few recent exceptions
\cite{coviello2012,jia2014network,zhou2014bidirectional}) concentrate
on systems without topology, also known as fully-connected systems,
in which one connects to all others in the other party and competes
with all others in the same party. In reality, however, one knows
only a limited number of others as characterized by the degree distribution of the social network. Hence a simple but fundamental question arises: what is the outcome of the matching process when topology is present?

In this work, we aim to address this question by analyzing the
impact of network structure on the specific example of the process
of matching, namely, human mate choice. Our motivation
to address this question is caused not only by the limited knowledge
on this matter, but also by the fact that topology could
fundamentally change properties of the system and further affect its
dynamical process. We have witnessed evidence of such impact,
accumulated in the last decades from the advances towards
understanding complex networks: a few shortcuts on a regular lattice
can drastically reduce the mean separation between nodes and give
rise to the small-world phenomenon \cite{Watts-Nature-98,
jia2013structural}, the power-law degree distribution of scale-free
networks can eliminate the epidemic threshold of epidemic spreading
\cite{pastor2001epidemic, boguna2003absence} and synchronization can
be reached faster in networks than in regular lattices
\cite{lago2000fast, wang2002synchronization,
nishikawa2003heterogeneity}. Indeed, numerous discoveries have been
made in different areas when considering topology in the analysis of
many classical problems \cite{Liu-Nature-11, jia-naturecomm-2013,
yuan2013exact, jia2014connecting, gao2014target, watts2002simple,
singh2013threshold, lu2008naming, xie2011social,
thompson2014propensity}. Hence it is fair to expect that the network
topology would also bring new insights on the matching process that
we are interested in.

% You may title this section "Methods" or "Models". 
% "Models" is not a valid title for PLoS ONE authors. However, PLoS ONE
% authors may use "Analysis" 
\section*{Methods}

\begin{figure}[h]
\begin{center}
\resizebox{10cm}{!}{\includegraphics{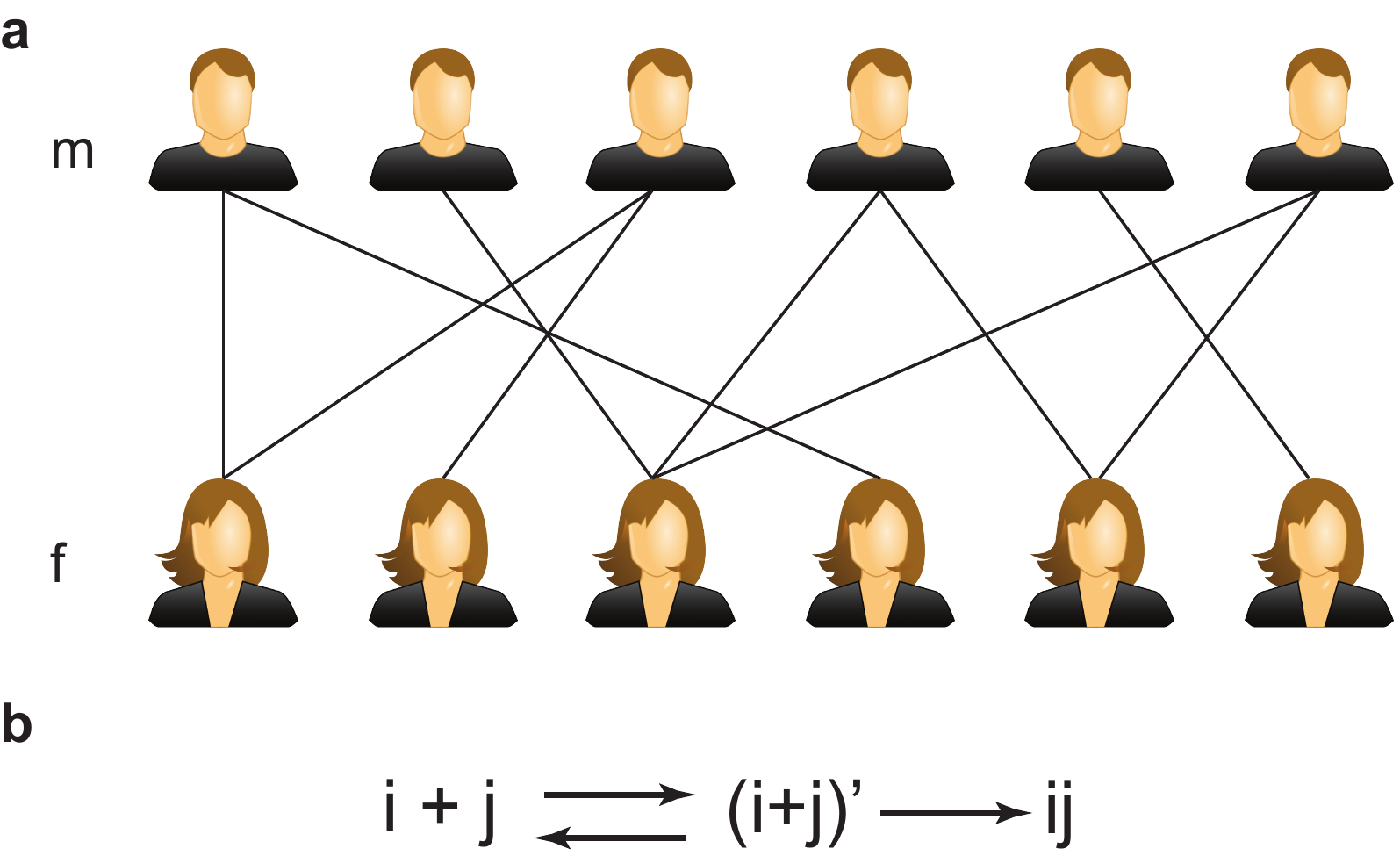}}
\caption{{\bf (a)} An example of a bipartite graph, which is composed of two disjoint sets of nodes $m$ and $f$. There is no link between nodes in the same set and the connection between sets is characterized by degree distribution $P(k)$. {\bf (b)} The action scheme of the mate choosing process. Two nodes $i$ and $j$ have to undergo an intermediate stage to reach the stable long term relation. During the intermediate stage nodes $i$ and $j$ are also available to build relationship with other nodes. If this happens they break and their relationship is back to the initial state. \label{fig:figure1}}
\end{center}
\end{figure}

We start with a bipartite graph with $2N$ nodes. The bipartite graph consists of two disjoint sets $m$ and $f$ of equal size, representing two parties, each with $N$ members.  While our model can be more general, for simplicity, we consider the two parties as collections of heterosexual male and female individuals (Fig. \ref{fig:figure1}.a). Each node, representing one individual, has $k$ links drawn from the degree distribution $P(k)$, randomly connecting to $k$ nodes in the other set. On average, a node has $\la k \ra = \sum kP(k)$ links, referred to the average degree of the network. To characterize the process of human mate choice, each node is assigned a random number $a$ as its attractiveness drawn uniformly from the range $[0,1)$.  Combining features in some previous works \cite{kalick1986matching, simao2003emergent} with the network structure, we consider the process of human mate choice as a two-step stochastic process which generates the numerical model as follows (Fig. \ref{fig:figure1}.b):

1. At each discrete time step, randomly pick a link. Let's denote the nodes connected by this link as node $i$ and node $j$ and their attractiveness as $a_i$ and $a_j$, respectively.

2. Draw two random numbers independently and uniformly
from the range $[0,1)$, denoted by $r_i$ and $r_j$. Check the
matching condition defined as $a_i > r_j$ and $a_j > r_i$.

3. If the matching condition is satisfied and nodes $i$ and $j$ are
not in a relationship with each other, pair them into intermediate
pairing and dissolve them from any previous intermediate pairing
with other nodes, if there are any.

4. If the matching condition is satisfied and nodes $i$ and $j$ are already in the intermediate pairing with each other, join them into the stable couple. Make nodes $i$ and $j$ unavailable to others by removing them from the network together with all their links.

5. Repeat from step 1 until there is no link left.

The matching condition in step 2 ensures that individuals
mutually accept each other.
%%%%%%%%%%%%%%%%%%%%%%%%%%%%%%%%%%%%%%%%%%%%%%%%%%%%%%%%%%%%%%%%%%%%%%
The decision making is probabilistic: the probability
that node $i$ accepts node $j$ is $a_j$ (independent of its
own attractiveness $a_i$).
%%%%%%%%%%%%%%%%%%%%%%%%%%%%%%%%%%%%%%%%%%%%%%%%%%%%%%%%%%%%%%%%%%%%%%
A pairing is successfully established only when both
individuals decide to accept each other. The intermediate pairing
created in step 3 corresponds to the tendency of people not to fully
commit to a relationship at the beginning and to form a stable
couple only after such unstable intermediate stage.
The removal of nodes and links in step 4 merely accelerates the
simulation, as these links should not be considered by others and the
corresponding nodes in the stable state are not available for
matching. Undoubtedly our model only captures a very small fraction
of features in the matching process. The goal of this work is
not to propose a sophisticated model that is able to regenerate all
observations in reality. Instead, we focus on attractiveness and
popularity (degree) that are essential in this process, hence this
model could be the simplest to study the interplay between these two
factors, shedding light on the effect of topology on this process.

To study the effects of topology, we focus on three most commonly used network structures with different degree distributions. 1) random k-regular graph (RRG) whose degree distribution follows a delta function $P(k) = \delta(k-\la k \ra)$, where $\la k \ra$ is the average degree of the network, corresponding to an extreme case that each person knows exactly the same number of others; 2) Erd\H{o}s-R\'enyi network (ER) with a Poisson degree distribution $P(k) = e^{-\la k \ra}\la k \ra^k/k!$, representing the situation that most nodes have similar number of neighbors and nodes with very high or low degrees are rare \cite{Erdos-PMIHAS-60}; 3) scale-free network (SF) generated via static model whose degree distribution has a fat-tail $P(k) \sim k^{-\gamma}$, featuring a large number of low degree nodes and few high degree hubs \cite{Barabasi-Science-99, Goh-PRL-01}. The constructions of these networks are as follows.

{\bf Constructing a random k-regular graph.} We start from two sets (sets $m$ and $f$) of $N$ disconnected nodes indexed by integer number $i$ ($i=1,\ldots N$). For each node $i$ in the set $m$, connect it to nodes $i$, $i+1$, $\ldots$ and $i+k-1$ in the set $f$ (using periodic boundary condition such that node $N$ in the set $m$ connects to node $N$, 1, $\ldots$ and $k-2$ in the set $f$, and so on). Then randomly pick two links, assuming that one link connects nodes $i$ in the set $m$ and $j$ in the set $f$ and the other connects nodes $i^\prime$ in the set $m$ and $j^\prime$ in the set $f$. Check if there is a connection between nodes $i$ and $j^\prime$ and nodes $i^\prime$ and $j$. If not, remove original links and connect nodes $i$ and $j^\prime$ and nodes $i^\prime$ and $j$. Repeat this process sufficiently large number of times such that connections of the network are randomized.

{\bf Constructing an Erd\H{o}s-R\'enyi network.} We start from two sets (sets $m$ and $f$) of $N$ disconnected nodes indexed by integer number $i$ ($i=1,\ldots N$). Randomly select two nodes $i$ and $j$ respectively from sets $m$ and $f$. Connect nodes $i$ and $j$ if there is no connection between them. Repeat the procedure until $N \la k\ra$ links are created.

{\bf Constructing a scale free network.} The scale-free networks analyzed are generated via the static model. We start from two sets (sets $m$ and $f$) of $N$ disconnected nodes indexed by integer number $i$ ($i=1,\ldots N$). The weight $w_i = i^{-\alpha}$ is assigned to each node, where $\alpha$ is a real number in the range $[0,1)$. Randomly selected two nodes $i$ and $j$ respectively from sets $m$ and $f$, with probability proportional to $w_i$ and $w_j$. Connect nodes $i$ and $j$ if there is no connection between them. Repeat the procedure until $N \la k\ra$ links are created. The degree distribution under this construction is $P(k) = \frac{[\la k \ra (1-\alpha)/2]^{1/ \alpha}}{\alpha}
\frac{\Gamma(k-1/ \alpha, \la k \ra (1-\alpha)/2)}{\Gamma(k+1)}$ where $\Gamma(s)$ the gamma function and $\Gamma(s,x)$ the upper incomplete gamma function. In the large $k$ limit, the distribution becomes $P(k) \sim k^{-(1+\frac{1}{\alpha})} = k^{-\gamma}$.

{\bf Introducing correlations between the attractiveness and the degree.} We generate $2N$ random numbers drawn between 0 and 1 and sort them in ascending order and index them by integer number $i$ ($i=1,\ldots 2N$). We sort nodes of networks in ascending order of their degrees and index them by integer number $j$ ($j=1,\ldots 2N$). For positive correlation between the degree and attractiveness, assign $i^{th}$ random number as the attractiveness of node $j=i$. For negative correlation between the degree and attractiveness, assign $i^{th}$ random number as the attractiveness of node $j=2N-i+1$.

% Results and Discussion can be combined.
\section*{Results}
\subsection*{Effects of Network Topology on the Correlation in Attractiveness}

The matching hypothesis suggests similarities in attractiveness between the two coupled individuals. To test it, we employ the Pearson coefficient of correlation $\rho$ as a measure of similarity, that is defined as
\begin{equation}
\rho = \frac{\sum_{i}^{n}(a_{m,i} - \overline{a}_m)(a_{f,i} - \overline{a}_{f})}{\sqrt{\sum_{i}^{n}(a_{m,i} - \overline{a}_m)^2}\sqrt{\sum_{i}^{n}(a_{f,i} - \overline{a}_f)^2}},
\end{equation}
where $a_{m,i}$ and $a_{f,i}$ are the attractiveness of the individuals in sets $m$ and $f$ of the $i^{\text{th}}$ couple, $\overline{a}_m$ and $\overline{a}_f$ are the average attractiveness of the matched individuals in sets $m$ and $f$ and $n$ is the number of matched couples in the network. The Pearson coefficient of correlation $\rho$ varies from -1 to 1, where 1 corresponds to the strongest positive correlation when two quantities are perfectly linearly increasing with each other, whereas -1 is the strongest negative correlation when two quantities are perfectly linearly dependent and one decreases when the other increases.

\begin{figure}[h]
\begin{center}
\resizebox{12cm}{!}{\includegraphics{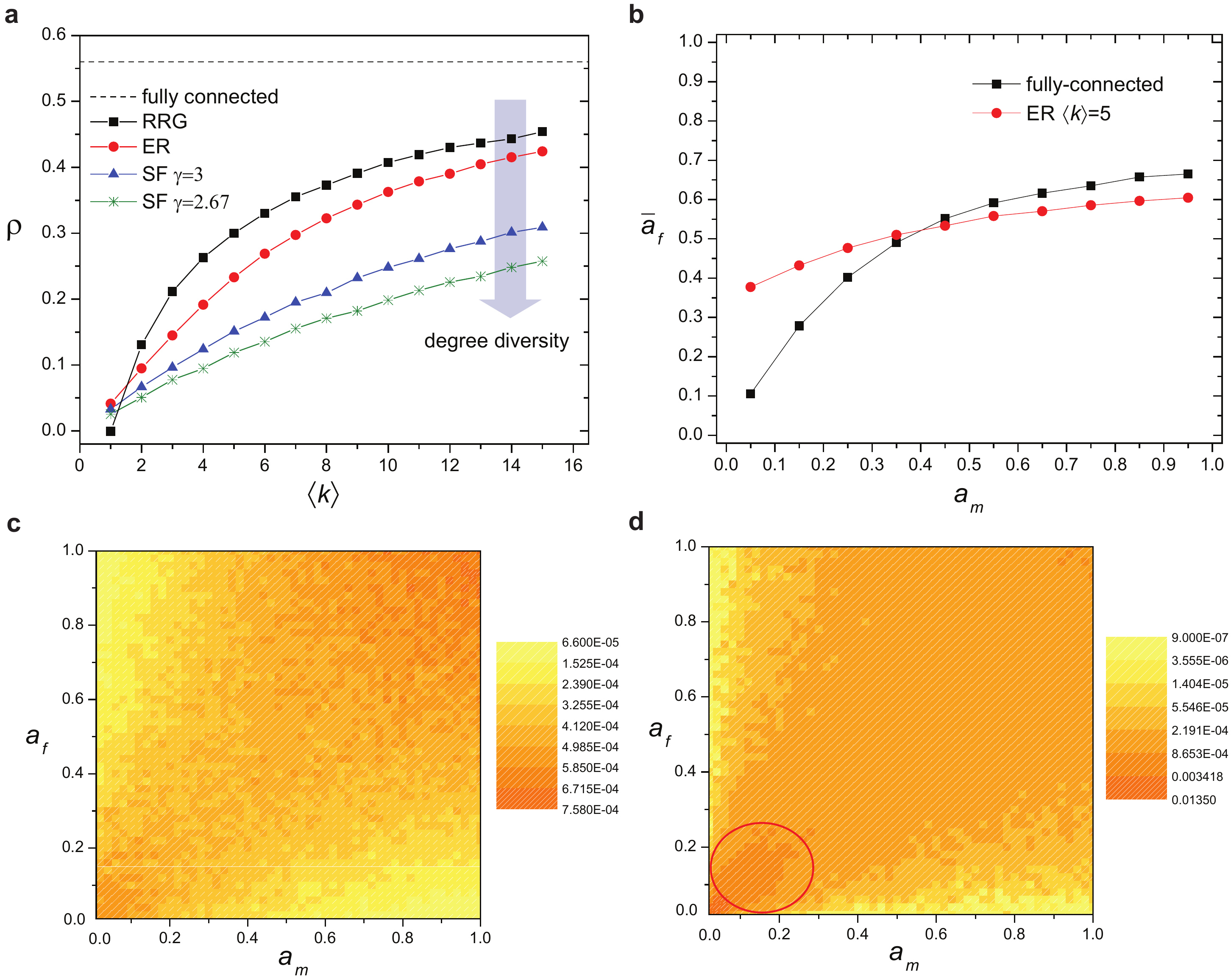}}
\caption{{\bf (a)} The Pearson coefficient of correlation $\rho$ of the attractiveness between the two coupled individuals in different systems. $\rho$ is strongest in fully-connected systems. In sparse networks, $\rho$ increases monotonically with the average degree $\la k \ra$ and decreases with the degree diversity. For all cases investigated, system size is $2N$ and $N = 10,000$.
{\bf (b)} The average attractiveness $\overline{a}_f$ of individuals in the set $f$ who are matched with those in a subset of $m$ with attractiveness in the range $[a_m - 0.05, a_m + 0.05)$ for a series of points $a_m$. In fully-connect systems, the less attractive individuals are bound to be coupled with ones who are also less attractive. In sparse networks, however, they are coupled with ones who are more attractive. {\bf (c)} The attractiveness contour figure of the coupled individuals in Erd\H{o}s-R\'enyi networks with average degree $\la k \ra = 5$. A pattern emerges even when similarity is not the motivation in seeking partners. $a_m$ and $a_f$ are the attractiveness of nodes in sets $m$ and $f$, respectively. {\bf (d)} The attractiveness contour figure of the coupled individuals in fully-connected systems. The correlation is strongest towards the less attractive individuals (the circled part).  \label{fig:figure2}}
\end{center}
\end{figure}

We first check the scenario studied in most of the previous works, when topology is not considered and each node is potentially able to match an arbitrary node in the other set. Our model generates a high correlation of the couple's attractiveness with the average $\rho \approx 0.56$ (Fig. \ref{fig:figure2}.a). This value is similar to the result generated in the previously proposed model which accounts also for attractiveness decay \cite{kalick1986matching} even though this feature is not present in ours. It is noteworthy that similarity is not explicitly considered when establishing a matching in this model and each individual only seeks attractive partners. However, the mutual agreement between two individuals effectively depends on the joint attractiveness of both. Hence individuals with high attractiveness will have the advantage in finding highly attractive partners, causing them to be removed from the dynamics soon, while less attractive individuals find their matches later. Therefore, as time goes on, only less and less attractive individuals are available to form a couple, thus they are more likely to get a partner with similar attractiveness.

The positive correlations in attractiveness are also observed in all three classes of networks studied. They are lower than the correlation observed in the fully-connected systems but increase monotonically with the average degree $\la k \ra$. Furthermore, as the network degree distribution varies from a delta function to a Poisson distribution and to a fat-tail distribution, the variance in the degree distribution increases. Our results indicated that for a given $\la k \ra$, $\rho$ decreases with the increased degree diversity (Fig. \ref{fig:figure2}.a). In other words, the broader the degree distribution is, the lower the correlation in attractiveness between the two coupled individuals will be. The reason is that as the degree diversity increases, more and more links are connected to a few high degree nodes. The majority of nodes have lower degrees compared to the network with the same degree but smaller degree diversity. Hence the majority of nodes have less opportunities in selecting partners and therefore smaller chance to find a partner with closely matched attractiveness. As the result the attractiveness correlation decreases.

While the correlation in attractiveness is strongest when the system is fully-connected, we find that the difference in the correlations is caused mostly by the matched individuals with low attractiveness. Indeed, the average attractiveness of those who are coupled with highly desired individuals does not depend much on the presence of the network structure (Fig. \ref{fig:figure2}.b-d).
In fully-connected systems, less attractive individuals are bound to be coupled with partners of low attractiveness, which contributes significantly to the total correlation $\rho$. In sparse networks, however, if they successfully find partners, their partners are likely to be more attractive than them. Therefore, the limited choice in sparse networks reduces competitions among individuals, especially for those with low attractiveness, hence giving rise to lower attractiveness correlations between the two coupled individuals.

\begin{figure}[h]
\begin{center}
\resizebox{12cm}{!}{\includegraphics{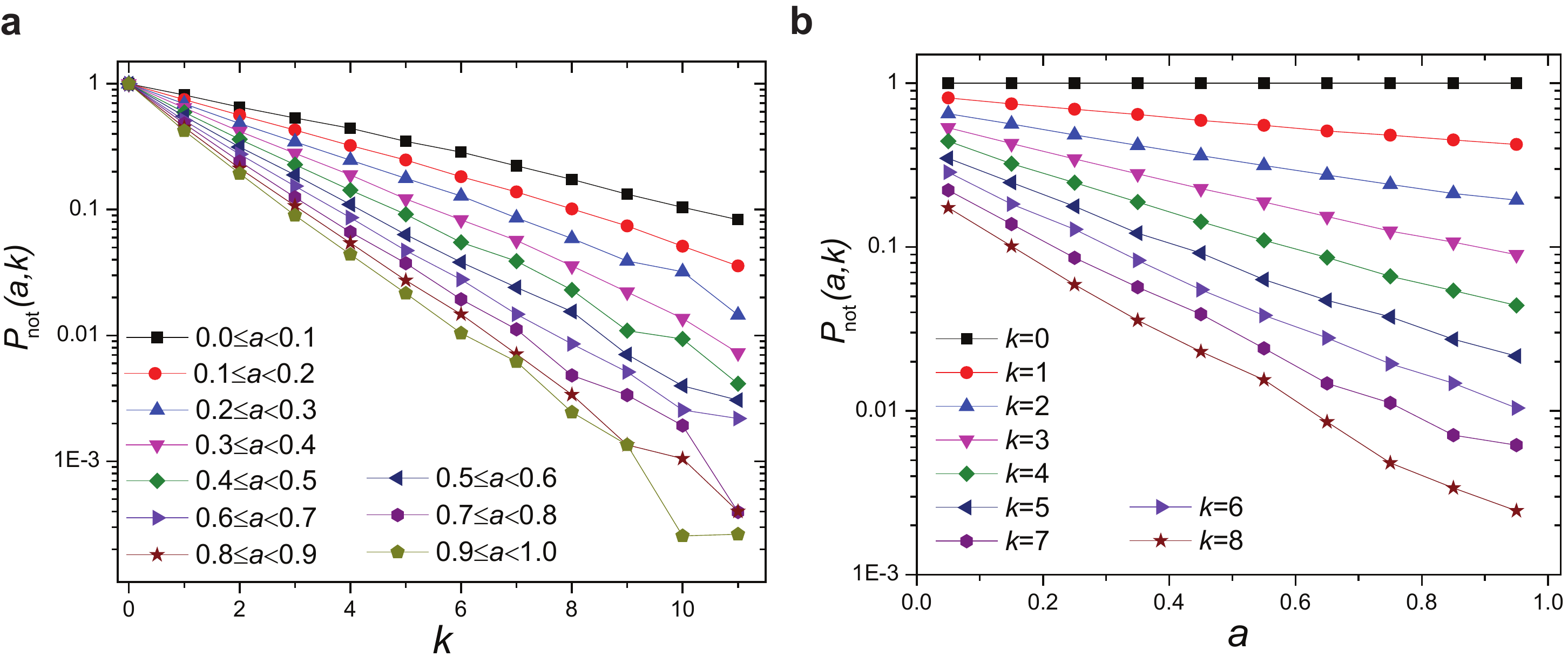}}
\caption{$\bf{(a, b)}$ The probability of failing to be matched conditioned on attractiveness $a$ and degree $k$ ($P_\text{not}(a,k)$) decreases exponentially with $a$ and $k$ in scale-free networks with $P(k) \sim k^{-\gamma}$, $\gamma = 3$ and $\la k \ra = 5$. \label{fig:figure3}}
\end{center}
\end{figure}

In fully-connected systems all individuals are able to find their partners. But in networks one faces a chance of failing to be matched. How often it occurs depends on one's popularity (degree) and attractiveness. Here we consider $P_\text{not}(a,k)$ defined as the probability of failing to be matched conditioned on degree $k$ and attractiveness within the range $[a-0.05, a+0.05)$. We find that $P_\text{not}(a,k)$ drops exponentially with both degree $k$ and attractiveness $a$. This implies that getting more popular brings the similar benefit as being more attractive in terms of finding a partner (Fig. \ref{fig:figure3}).

\begin{figure}[h]
\begin{center}
\resizebox{7cm}{!}{\includegraphics{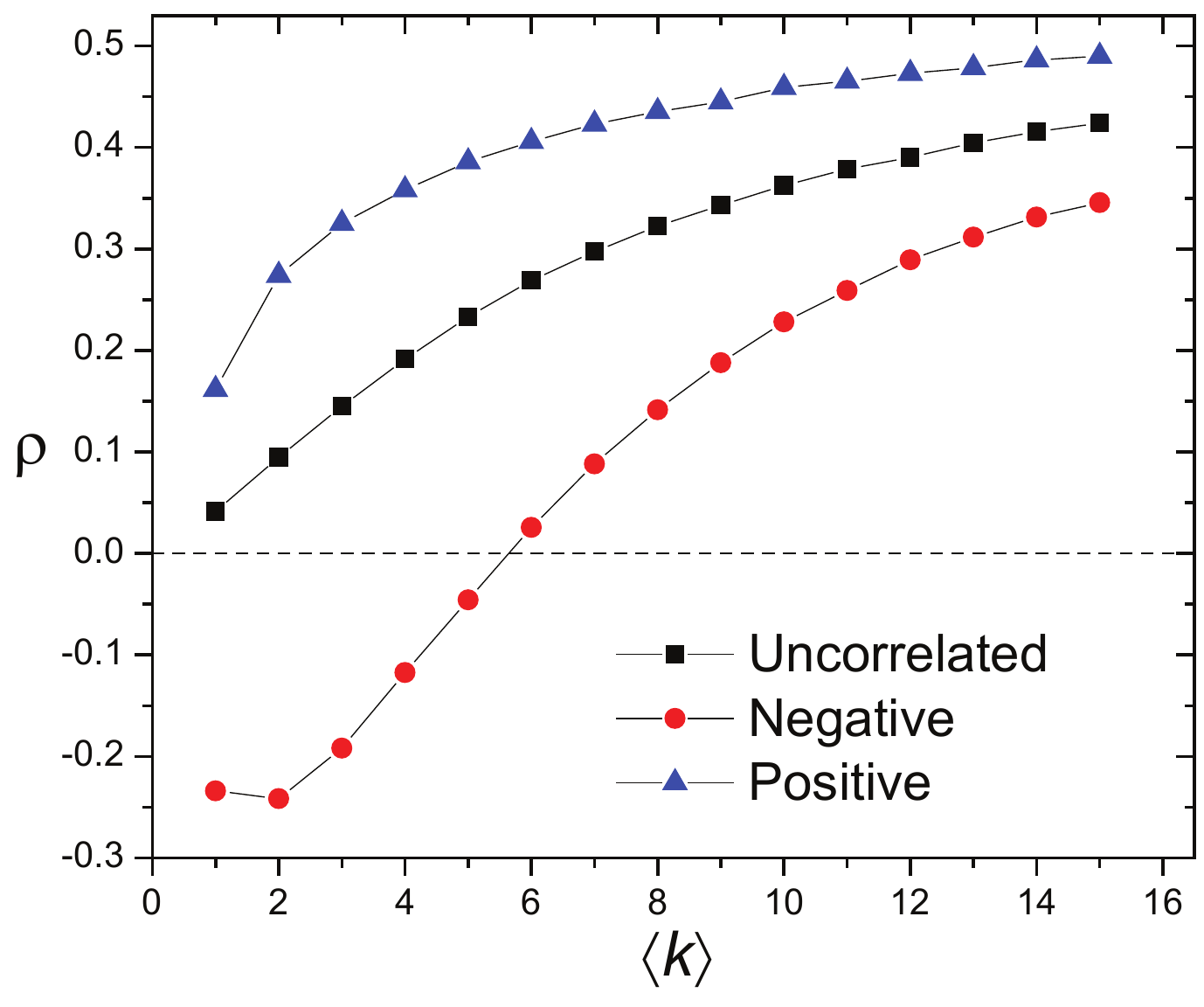}}
\caption{The Pearson coefficient of correlation $\rho$ of the attractiveness between the two coupled individuals in Erd\H{o}s-R\'enyi networks with size $2N$ ($N=10,000$) and varying average degree $\la k \ra$. $\rho$ increases monotonically in all three cases analyzed. However, $\rho$ is largest in networks where the degree and the attractiveness are positively correlated. When they are negatively correlated, $\rho$ is weakest and can even be negative. \label{fig:figure4}}
\end{center}
\end{figure}

So far we have concentrated only on cases where there is no correlation between one's popularity (degree) and attractiveness. In reality these two features are often correlated. On one hand, the positive correlation is somewhat expected as a highly attractive person can potentially be also very popular hence having a larger degree. On the other hand, negative correlation could also occur when those with low attractiveness are more active in making friends to balance their disadvantage in attractiveness. We extend our analysis to two extreme cases when degree and attractiveness are correlated (see Method). For a given network topology, the correlation of attractiveness ($\rho$) is strongest when the degree and the attractiveness are positively correlated and weakest when they are negatively correlated. It is noteworthy that with negative degree-attractiveness correlation, $\rho$ can become negative in networks with low $\la k \ra$, suggesting that the matching hypothesis may not hold in such networks even though the underlying mechanism does not change (Fig. \ref{fig:figure4}).

% Please do not create a heading level below \subsection. For 3rd level headings, use \paragraph{}.
\subsection*{Number of Couples Matched}

\begin{figure}[h]
\begin{center}
\resizebox{12cm}{!}{\includegraphics{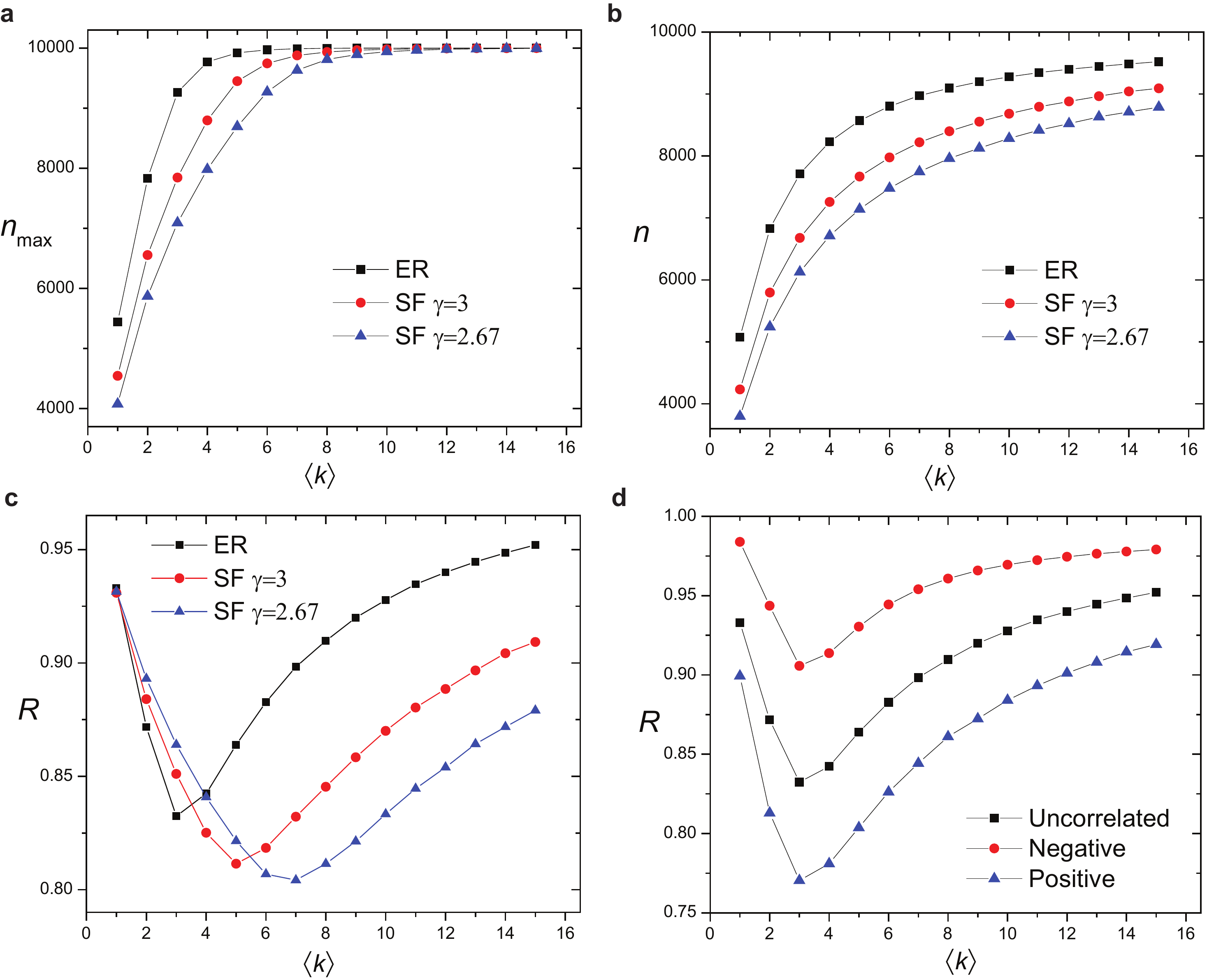}}
\caption{{\bf (a)} The size of the maximum matching $n_\text{max}$ increases monotonically with the average degree $\la k \ra$ in different networks. {\bf (b)} The number of matched couples $n$ increases monotonically with the average degree $\la k \ra$ in different networks. {\bf (c)} The ratio between the number of matched couples and the size of the maximum matching ($R = n/n_\text{max}$) varies non-monotonically with the average degree $\la k \ra$. {\bf (d)} Different behaviors of $R$ in Erd\H{o}s-R\'enyi networks where the correlation between degree and the attractiveness varies. Negative correlation between the degree and the attractiveness yields the largest $R$ while positive correlation between the degree and the attractiveness results in the smallest $R$. Networks tested in all cases are with size $2N$ ($N=10,000$). \label{fig:figure5}}
\end{center}
\end{figure}

Another quantity affected by
topology and typically studied is the number of couples a system can
eventually match $n$ \cite{zhou2014model, zhou2014bidirectional}. When
the system is fully-connected, everyone can find a partner and the
number of couples is  $n=N$. In sparse networks, typically there are 
fewer matched couples than $N$ and the highest number of matched couples $n_\text{max}$ is given by the maximum matching which disregards the attractiveness \cite{Karp-IEEE-81, Hopcroft-SIAM-73}.
To measure the performance of the system in terms of the matching,
we focus on the quantity $R=n/n_\text{max}$ defined as the {\em ratio} between the
number of couples matched and the size of the maximum matching. While both
the number of the couples matched and the size of the maximum matching increase monotonically as
the network becomes denser (Figs. \ref{fig:figure5}.a, b), their ratio $R$ changes
non-monotonically with $\la k \ra$ (Fig. \ref{fig:figure5}.c). The
system's performance can be relatively good when the network is very sparse
or very dense, but relatively poor for the intermediate range of density. This
is mainly because when more links are added to the system, the
number of couples matched increases slower than the size of the maximum matching;
only when this size becomes saturated to $N$ the ratio $R$
starts to increase with $\la k \ra$.

Correlation between the degree and attractiveness also plays a role
in the value of $R$ achieved by a network. The maximum matching $n_\text{max}$ depends only
on the topology of the network and does not depend on the attractiveness. A successful matching between two
nodes in our model, however, depends on both their attractiveness and
their degrees. Therefore, $R$ depends on the degree-attractiveness
correlation. In both cases when either positive or negative correlation
between degree and attractiveness is present, $R$ varies
non-monotonically with $\la k \ra$ just like in the case when there is
no degree-attractiveness correlation (Fig. \ref{fig:figure5}.d).
However, negative correlation between degree and attractiveness
yields more while positive correlation yields fewer matched couples than that when degree and
attractiveness are uncorrelated. Considering the fact that the
similarity between the two coupled individuals ($\rho$) is largest in networks with
positive degree-attractiveness correlation and smallest with
negative degree-attractiveness correlation, such a dependence of $R$ on degree-attractiveness correlation
implies that the system's performance in terms of the number of matched couples is better when it is less
selective.

\section*{Discussion}

In summary, we studied the effect of topology on the process of
human mate choice. In general, our findings support the conclusion
of the previous works that similarities in attractiveness between
coupled individuals occur even though the similarity is not the
primary consideration in searching for partners and each individual only seeks attractive partners, in agreement with
the matching hypothesis. When topology is present, the extent of
such similarity, measured by Pearson coefficient of correlation,
grows monotonically with the increased average degree and
decreased degree diversity of the network. The correlation is
weaker in sparse networks because in them the less attractive
individuals who are successful in finding partners, are likely to be
coupled with more attractive mates. In fully-connected systems,
however, they are almost certain to be coupled with partners also
less attractive, contributing significantly to the total attractiveness correlation.

Another effect of the topology is that one faces a chance of
failing to find a partner. Such the chance decays exponentially with
one's attractiveness and degree, therefore being more popular can
bring benefits in terms of finding a partner similar to being more
attractive. The correlation of couple's attractiveness is also
affected by the degree-attractiveness correlation, which is
strongest in networks where attractiveness and popularity are
positively correlated and weakest when they are negatively
correlated. In networks with negative degree-attractiveness
correlation, the attractiveness correlation between coupled individuals can
be negative when the average degree is low, implying that matching
hypothesis may not hold in such systems. Finally, the number of
couples matched also depends on the topology. The ratio between the
number of matched couples and the maximum number of couples that can be matched, denoted as $R$,
changes non-monotonically with the average degree. $R$ is largest in
networks with negative degree-attractiveness correlation and
smallest when the attractiveness and the popularity are positively
correlated.

%%%%%%%%%%%%%%%%%%%%%%%%%%%%%%%%%%%%%%%%%%%%%%%%%%%%%%%%%%%%%%%%%%%%%%%%%%
The non-monotonic behavior of the matching ratio $R$ is also
interesting from a stochastic optimization viewpoint: the simple
trial-and-error matching process, governed and constrained by
individuals' attractiveness, fares reasonably well everywhere
(against the maximum attainable matching on a given bipartite
graph), except for a narrow intermediate sparse region
(Fig.~\ref{fig:figure5}). The ``worst-case" average degree depends
strongly on network heterogeneity but {\em not} on
degree-attractiveness correlations.
%%%%%%%%%%%%%%%%%%%%%%%%%%%%%%%%%%%%%%%%%%%%%%%%%%%%%%%%%%%%%%%%%%%%%%%%%%%

Our results revealed the role of topology in the process of human
mate choice and can bring further insights into the investigations
of different matching processes in different networks
\cite{laureti2003matching, wang2007impact, lage2006marriage,
zhou2014model, zhou2014bidirectional}.
Indeed, in this work we focused only on the basic model of the mate seeking
process in random networks. However, different variations can be considered.
For example, there is no degree correlation between the two coupled individuals
observed in our model, simply because the networks we studied are
random with no assortativity. In reality, the connection may not be
random and then assortativity can be considered. Furthermore, the networks in
our model are static and the degree of a node does not change with
time. In reality, a node may gain or lose friends and consequently
its degree may change. Likewise, stable matching between individuals does not have to last forever, it just needs to be an order of magnitude longer than unstable matching. It is possible to establish certain rates to stable matching dissolution and analyze the steady state behavior of so generalized system. Finally, here we considered the
attractiveness as a one dimensional attribute of individuals. In
more realistic scenarios, attractiveness can be a multi-dimensional
variable with different merits \cite{buss1994strategies,
ramsey2011mutual, he2013potentials}. Investigations of such more complicated cases are left to future work.

%\section*{Supporting Information}

% Include only the SI item label in the subsection heading. Use the \nameref{label} command to cite SI items in the text.
%\subsection*{S1 Video}
%\label{S1_Video}
%{\bf Bold the first sentence.}  Maecenas convallis mauris sit amet sem ultrices gravida. Etiam eget sapien nibh. Sed ac ipsum eget enim egestas ullamcorper nec euismod ligula. Curabitur fringilla pulvinar lectus consectetur pellentesque.

% Do NOT remove this, even if you are not including acknowledgments.
\section*{Acknowledgments}
This work was supported in part by the Army Research Laboratory under Cooperative Agreement Number W911NF-09-2-0053 and by the Office of Naval Research (ONR) Grant No. N00014-09-1-0607. The views and conclusions contained in this document are those of the authors and should not be interpreted as representing the official policies, either expressed or implied, of the Army Research Laboratory or the U.S. Government. 

%\nolinenumbers

%\section*{References}

%\bibliographystyle{plos2009}
%\bibliography{D:/Dropbox/Latex/BIB/matching}
% Either type in your references using
% \begin{thebibliography}{}
% \bibitem{}
% Text
% \end{thebibliography}
%
% OR
%
% Compile your BiBTeX database using our plos2009.bst
% style file and paste the contents of your .bbl file
% here.
% 

\newpage

\noindent

\newpage

\noindent

\newpage

\noindent

\newpage

\noindent

\newpage

\noindent

\end{document}